\documentclass[10pt,a4paper]{article}

\usepackage[latin1]{inputenc}
\usepackage{amsmath}
\usepackage{amsfonts}
\usepackage{amssymb}
\usepackage{moreverb, bm, booktabs}
\usepackage{graphicx}
\usepackage[affil-it]{authblk}
\usepackage[dvips,colorlinks,bookmarksopen,bookmarksnumbered,citecolor=red,urlcolor=red]{hyperref}

\providecommand{\keywords}[1]{\textbf{Keywords:} #1}
\newcommand{\rmd}{\mathrm{d}}
 \DeclareBoldMathCommand \y {y} 
\begin{document}
\title{Overcoming computational inability to predict clinical outcome from high-dimensional patient data using Bayesian methods}
       \date{}
\author{A.\,SHALABI \quad A.\,C.\,C.\,COOLEN
\vspace*{-3mm}
	\and Institute for Mathematical and Molecular Biomedicine
	\vspace*{-4.2mm}
        \and King's College London
        \vspace*{-4.2mm}
        \and London, U.K
        \vspace*{-4.2mm}
        \and akram.shalabi@kcl.ac.uk
        \vspace*{-4.2mm}
        \and ton.coolen@kcl.ac.uk
        \and
        E.\,de\,RINALDIS
        \vspace*{-3mm}
	\and NIHR Biomedical Research Centre
	\vspace*{-4.2mm}
        \and  R\&D Department
        \vspace*{-4.2mm}
        \and Guy`s Hospital
        \vspace*{-4.2mm}
        \and London, U.K.
        \vspace*{-4.2mm}
        \and  emanuele.de\_rinaldis@kcl.ac.uk
        }

            \maketitle

\begin{abstract}
Clinical outcome prediction from high-dimensional data is problematic in the common setting where there is only a relatively small number of samples. The imbalance causes data overfitting, and outcome prediction becomes computationally expensive or even impossible. 
We propose a Bayesian outcome prediction method that can be applied to  data of arbitrary dimension $d$, from $2$ outcome classes, and reduces overfitting without any approximations at parameter level. This is achieved by avoiding numerical integration or approximation, and solving the Bayesian integrals \textit{analytically}. We thereby reduce the dimension of numerical integrals from $2d$ dimensions to $4$, for any $d$. For large $d$, this is reduced further to $3$, and we obtain a simple outcome prediction formula without integrals in leading order for very large $d$. We compare our method to the \textit{mclustDA} method (Fraley and Raftery 2002), using simulated and real data sets. Our method perform as well as or better than \textit{mclustDA} in low dimensions $d$. In large dimensions $d$, \textit{mclustDA} breaks down due to computational limitations, while our method provides a feasible and computationally efficient alternative.
\end{abstract}

\hspace*{2.3mm} \keywords{Discriminant analysis; Bayesian outcome prediction; Overfitting; Curse of dimensionality; \\ \hspace*{27.9mm} Bayesian integration in high dimensions; Binary-class prediction.}


\section{Introduction}
Outcome prediction is based on discriminant analysis,
which is the use of known classifications of some observations to classify others \cite{Hastie1996}.  The aim is to use some observations in the data set under study as the training set, to automatically discover and learn the rules linking an observation to its class. These rules can then be applied to the remaining observations in the test or validation set to assign them to classes. Discriminant analysis methods are often probabilistic, or model-based, where the observations in each class are assumed to be generated by a distribution specific to that class  \cite{Hastie1996}. 

Fraley and Raftery {\cite{Fraley2002} developed a popular model-based discriminative analysis approach to outcome prediction that has been used in a variety of contexts \cite{Dean2006, Fraley2007, Iverson2009, Murphy2010, Andrews2012}. This involves producing Bayesian posterior probabilities of belonging to a class, or class-conditional probabilities, by using model-based clustering to fit a Gaussian mixture model (GMM) to each class in the training set, and estimating the optimal parameters 
whilst ignoring the uncertainty in these estimates. These are then  used to calculate the class-conditional probabilities of the remaining observations.

Using leave-one-out cross-validation (LOOCV) \cite{Lachenbruch1968, Stone1974, Meek2001}, we found that whilst their approach produces accurate and efficient results for data sets consisting of a limited number of samples with low dimensionality, a number of challenges arise for data sets with large dimensions. These are data overfitting \cite{Clarke2008, Michiels2011}, and computational inability \cite{Clarke2008}. These factors combined produce outcome predictions of limited reliability if any at all. Each parameter contributes a $d$-dimensional Bayesian integral to the predictive distribution, which is at best computationally expensive to estimate numerically. Instead, at the expense of ignoring the uncertainty in the parameters, they are estimated by the most probable values using the Expectation-Maximisation algorithm (EM) \cite{Dempster1977, Fraley2002}. Yet this is also computationally expensive in high-dimensions \cite{Fraley2002, Bouveyron2007, McNicholas2011}. Several dimension-reduction approaches such as principal component analysis (PCA) \cite{Joliffe2002, Fraley2002, Mclustversion4}, subspace clustering \cite{Scott1983, Bouveyron2007}, and/or using constrained and parsimonious models \cite{McNicholas2011}, have been suggested in literature. The first approach runs the risk of compromising the prediction accuracy through loss of information found in higher dimensions \cite{Chang1983, Bouveyron2014}. The second assumes that high-dimensional data live in subspaces with dimensionality less than $d$ \cite{Scott1983, Bouveyron2007, Bishop2006} but relies on finding a good classifier in high dimensions \cite{Bishop2006, Bouveyron2014}, and the third approach is a trade-off between perfect modelling and what can be correctly estimated in practice. The latter two have been combined to reduce the number of parameters and dimensions for EM to consider \cite{Bouveyron2007, Berge2011, Bouveyron2014}. However underestimated parameter uncertainty remains.

The solution to these challenges will involve some form of dimension reduction without underestimating parameter uncertainty or loss of information in the data. In this paper we show that it is possible to do so by solving the $d$-dimensional Bayesian integrals analytically, and propose a Bayesian outcome prediction method that can be applied to any arbitrary number of dimensions $d$ without any approximations at parameter level. In Section $2$, we briefly describe the method developed by Fraley and Raftery \cite{Fraley2002}, and in Section 3 we describe our Bayesian outcome prediction method. In Section $4$, we apply our method to simulated examples, and show how the curse of dimensionality \cite{Bellman1957} is lifted in terms of CPU and memory. In Section $5$, we apply our method to a real biomedical example. We conclude with a discussion in Section $6$.


\section{Model-based discriminant analysis}

Fraley and Raftery \cite{Fraley2002} developed a model-based approach to discriminant analysis which is  implemented in the  R software environment \cite{R2013} using the reference package \textit{mclust} \cite{Fraley2002, Mclustversion4}. In this section, we briefly describe their method. 

Many methods have been proposed for discriminant analysis, and it is applicable in a wide variety of settings \cite{Ripley1996, Duda2001, Mclachlan2003}. Discriminant analysis methods are often probabilistic, based on the assumption that observations in each class are generated by a distribution specific to that class. The number of classes $ J$ is assumed to be known. We typically have a multi-dimensional data set $D$ consisting of $N$ observations, and their corresponding class labels ${\sigma} \in \{0, \ldots, (J-1)\}$ that indicate which class each observation belongs to: i.e. $~D=\{({\bm x}_{1},\sigma_{1}),\ldots,({\bm x}_{N},\sigma_{N})\}$, consisting for each individual $i$, of a $d$-dimensional covariate vector ${\bm x}_{i}$, and the class $\sigma_{i}$ to which that individual belongs to. The data set is then split into a training and test set. Each pair $(\bm x_{i},\sigma_{i})$ is assumed to be is drawn independently, from an unknown joint probability distribution $p(\bm x,\sigma)$, that describes the data set being studied. In biomedical data sets, we typically  have $d \gg N$ \cite{Clarke2008, Michiels2011, Wang2008, Rinaldis2013, Naturebreast2012, Natureovarian2011, CCLE2012}. If we take $p(\bm x_{i}|\sigma)$ to be the probability distribution of the observations belonging to class $\sigma$, and $p(\sigma)$ as the proportion of observations that belong to that class, or class-imbalance parameter, then according to Bayes' theorem, the posterior probability that an observation $\bm x_{i}$ belongs to class $\sigma$, or class-conditional probability, is

\begin{equation}
p(\sigma|\bm x_{i}) = \frac{p(\bm x_{i}|\sigma)~p(\sigma)}{ \sum^{J-1}_{\sigma=0} p(\bm x_{i}|\sigma)~p(\sigma)}
\label{eqn:bayes_theorem}
\end{equation}

In probabilistic discriminant analysis, a model is fit to each class in the training set, and observations in the test set are assigned to the class corresponding to the model in which they have the highest class-conditional probability. Most common model-based discriminant analysis methods are based on the assumption that the observations in each class are multivariate Gaussian \cite{Dasgupta1998, Fraley1998}. GMMs are also commonly used \cite{Hastie1996, Fraley2002, McNicholas2011, Bouveyron2014}.

Fraley and Raftery \cite{Fraley2002} fit one GMM to each class in the training set using model-based clustering, and applied the optimal model to the test set. This withdrew the need to know the number of classes, and covariance matrix structure a priori by allowing the data in the training set to determine the number of classes in the data set. The procedure is called \textit{mclustDA} \cite{Fraley2002, Mclustversion4}, and can be summarised as follows: Set the number of GMM components, and type(s) of covariance matrix structure to consider. Apply model-based hierarchical clustering \cite{Fraley2002, Banfield1993, Dasgupta1998, Fraley1998} to the training set to provide initial, suboptimal partitions. These provide initial values for the parameters and class-conditional probabilities, which are used to initialise the Expectation-Maximisation algorithm (EM) \cite{Dempster1977, Fraley1998, Dasgupta1998}. EM
is applied to find the optimal parameters for each GMM, assigning a class-conditional probability to each data point. The Bayesian Information Criterion (BIC) \cite{Schwarz1978, Dasgupta1998, Fraley1998, Fraley2002} is then computed for each GMM, with the one with the highest BIC estimated to be the optimal model. Degenerate solutions are alleviated by using the posterior mode (MAP) instead of the maximum likelihood estimates (MLE) for the EM and BIC calculations \cite{Fraley2007}. The optimal model is then applied to the test set, and the class-conditional probabilities are calculated using (\ref{eqn:bayes_theorem}). An observation in the test set is assigned to the class corresponding to the highest class-conditional probability. It has been suggested that \textit{mclustDA} may not work well or be insufficient for high-dimensional data due to computational inability and/or overfitting \cite{Fraley2002, Wehrens2004, Fraley2005, McNicholas2011, Mclustversion4}. A suggested solution is to restrict the parameters of the model \cite{McNicholas2011, Mclustversion4} or apply a dimension-reduction technique such as PCA \cite{Mclustversion4}. We apply the former at the risk of fitting an inadequate model to the data rather than risk losing information in the data by using an a priori dimension reduction technique.

 We fit one isotropic Gaussian distribution to each class such that ${\bm x}|\sigma \sim \mathcal{N}({\bm \mu}_ {\sigma}, \lambda^{2}_ {\sigma} \, \mathbb{I}_{d\times d})$ where ${\bm \mu}_ {\sigma}$ is a $d \times 1$ mean vector, and $\lambda^{2}_ {\sigma}$ is the variance of the distribution for class $\sigma$. This is achieved by choosing \textbf{EII} (Equal (\textbf{E}) volume, Identity Matrix (\textbf{I}) shape and  Identity Matrix (\textbf{I}) orientation) as the covariance structure,  \textit{mclustDA} as the model type, \textit{defaultPrior} as the prior, and setting G (the number of Gaussians to be considered per class) to $1$ \cite{Fraley2002, Mclustversion4}.
Using LOOCV, we found that for data sets consisting of a limited number of samples ($N=100$) with $d=3, 10, 100, 1000$, \textit{mclustDA} produces predictions in a reasonably quick time with little computational expense. For data sets consisting of a large number of samples with low $d$, \textit{mclustDA} was computationally expensive due to the large amount of time taken by model-based hierarchical clustering \cite{Fraley2002, Wehrens2004, Fraley2005}. Taking a random subset of the data set as the training set, and classifying the remaining observations (in reasonably sized blocks) \cite{Banfield1993, Fraley1998, Fraley2002, Mclustversion4} is more efficient, but can lead to decreasing accuracy \cite{Meek2001, Wehrens2004, Fraley2005}. A more accurate incremental approach has been developed but only for data sets whose storage and memory requirements are not too large to be handled in whole by the \textit{mclust} software package \cite{Wehrens2004, Fraley2005}. 
For large $d$ (i.e. $d=10,000$), \textit{mclustDA} could not handle the large dimensionality of the data set, and failed to produce any results.
 We surmise that for a large number of samples, it would be extremely expensive in terms of CPU power and memory to produce any results given the combined computational expense of model-based hierarchical clustering, and EM. For a limited number of samples, we would be overfitting heavily. We describe a method that both overcomes this computational inability, and reduce overfitting without sacrificing any information contained in the data set in Section $3$.


\section{Bayesian binary outcome prediction for any number dimensions}

In this section, we present a Bayesian outcome prediction method that can be applied to $2$ homogeneous classes, and any number of dimensions. We reduce the dimension of numerical integrals from $2d$ dimensions to $4$, for any $d$ (FD). For large $d$, this is reduced further to $3$ (LD), and we obtain a simple outcome prediction formula without integrals in leading order for very large $d$ (LOLD).

We fit one isotropic Gaussian distribution to each class such that ${\bm x}|\sigma \sim \mathcal{N}({\bm \mu}_ {\sigma}, \lambda^{2}_ {\sigma}\, \mathbb{I}_{d\times d})$ where ${\bm \mu}_ {\sigma}$ is a $d \times 1$ mean vector, and $\lambda^{2}_ {\sigma}$ is the variance of the distribution for class $\sigma$. The parameter prior with respect to the mean is set to be ${\bm \mu}_ {\sigma} \sim \mathcal{N}(0, \alpha^{-1}_ {\sigma}\,\mathbb{I}_{d\times d})$, where $\alpha_ {\sigma}$ is the precision of the distribution, and use $\delta(\lambda_ {\sigma} - \psi_ {\sigma})$ to fix $\lambda_ {\sigma}$, and treat it as hyperparameter. We only take the uncertainty at parameter level into account. If we know precisely or wish to control the class-imbalance in the data set, we can fix $p(\sigma)$ to our chosen value $\sigma^{\prime}_ {\sigma}$. In the absence of any information about the class-imbalance in the data set, we take this to be the class-imbalance in the training set $\bar{\sigma}_ {\sigma} = {N_ {\sigma}}/{N}$ (the maximum entropy distribution for $p(\sigma)$), where $N_ {\sigma}$ is the number of observations in class $\sigma$ in the training set \cite{Bishop2006}. Proceeding with $\sigma=1$ ($p(0|{\bm x}^{\star}, D)$ can be calculated by subtracting $p(1|{\bm x}^{\star}, D)$ from $1$), the predictive distribution for a new observation ${\bm x}^{\star}$ is

\begin{align*}
p(1|{\bm x}^{\star}, D) &= \int^{\infty}_{-\infty} \int^{\infty}_{-\infty}\rmd {\bm \mu}_{0} \rmd {\bm \mu}_{1}\,p({\bm \mu}_{0}|D, \hat{\psi}_{0}, \hat{\alpha}_{0})\,p({\bm \mu}_{1}|D, \hat{\psi}_{1}, \hat{\alpha}_{1})\,\frac{p({\bm x}^{\star}|1, {\bm \mu}_{1}, \hat{\psi}_{1})\,\bar{\sigma}_{1}}{p({\bm x}^{\star}|0, {\bm \mu}_{0}, \hat{\psi}_{0})\,\bar{\sigma}_{0}\,+\,p({\bm x}^{\star}|1, {\bm \mu}_{1}, \hat{\psi}_{1})\,\bar{\sigma}_{1}}
\\
\\ 
&\propto \int^{\infty}_{-\infty}\int^{\infty}_{-\infty} \rmd {\bm \mu}_{0} \rmd {\bm \mu}_{1}\,\mathrm{e}^{-\frac{1}{2\hat{s}^{2}_{0}}({\bm \mu}_{0} -(\frac{\hat{s}_{0}}{\hat{\psi}_{0}})^{2}\sum_{i=1}^{N} \delta_{\sigma_{i}, 0} \,{\bm x}_{i})^{2} -\frac{1}{2\hat{s}^{2}_{1}}(\bm{\mu}_{1}-(\frac{\hat{s}_{1}}{\hat{\psi}_{1}})^{2} \sum_{i=1}^{N} \delta_{\sigma_{i}, 1} \,\bm x_{i})^{2}} \Big[1+\frac{p({\bm x}^{\star}|0, {\bm \mu}_{0}, \hat{\psi}_{0})\bar{\sigma}_{0}}{p({\bm x}^{\star}|1, {\bm \mu}_{1}, \hat{\psi}_{1})\bar{\sigma}_{1}}\Big]^{-1}
\\
\\
&\text{with} \quad \hat{s}^{-2}_{0} = \hat{\alpha}_{0} + \hat{\psi}^{-2}_{0} \sum^{N}_{i=1} \delta_{\sigma_{i}, 0}\quad\quad \hat{s}^{-2}_{1} = \hat{\alpha}_{1} + \hat{\psi}^{-2}_{1}  \sum^{N}_{i=1} \delta_{\sigma_{i}, 1}
\end{align*}
Transforming the integration variables using 
\begin{align*}
{\bm \mu}_{0} &= (\frac{\hat{s}_{0}}{\hat{\psi}_{0}})^{2} \sum_{i=1}^{N} \delta_{\sigma_{i}, 0} \,{\bm x}_{i} + \hat{s}_0 {\bm y_{0}}\quad \quad
{\bm \mu}_{1} = (\frac{\hat{s}_{1}}{\hat{\psi}_{1}})^{2} \sum_{i=1}^{N} \delta_{\sigma_{i}, 1} \,{\bm x}_{i} + \hat{s}_1 {\bm y_{1}}
\\
\\
&\text{where} \quad {\bm y_{0}} \sim \mathcal{N}(0,\mathbb{I}_{d\times d}) \quad \quad {\bm y_{1}} \sim \mathcal{N}(0,\mathbb{I}_{d\times d})
\end{align*}
 gives 
 \begin{align*}
 p(1|{\bm x}^{\star}, D) 
 &=  \frac{\bar{\sigma}_{1}\hat{\psi}^{d}_{0}}{\bar{\sigma}_{0}\hat{\psi}^{d}_{1}}\,\int^{\infty}_{-\infty}\int^{\infty}_{-\infty} \,
\Big[\frac{\bar{\sigma}_{1}\hat{\psi}^{d}_{0}}{\bar{\sigma}_{0}\hat{\psi}^{d}_{1}}+\mathrm{e}^{\frac{1}{2}(\frac{\hat{s}_1}{\hat{\psi}_{1}})^{2}\Big(\frac{{\bm x}^{\star}}{\hat{s}_{1}} - \frac{\hat{s}_{1}}{\hat{\psi}^{2}_{1}} \sum^{N}_{i=1}\delta_{\sigma_{i}, 1}~{\bm x}_{i} - {\bm y_{1}}\Big)^{2}- \frac{1}{2}(\frac{\hat{s}_0}{\hat{\psi}_{0}})^{2}\Big(\frac{{\bm x}^{\star}}{\hat{s}_{0}} - \frac{\hat{s}_{0}}{\hat{\psi}^{2}_{0}} \sum^{N}_{i=1}\delta_{\sigma_{i}, 0}~{\bm x}_{i} - {\bm y_{0}}\Big)^{2}}\Big]^{-1}
\\
&\quad\quad\quad\quad\quad\quad\quad\quad\quad\quad\quad\quad\quad\quad\quad\quad\quad\quad\quad\quad\quad\quad\quad\quad\quad\quad\quad\quad\quad\quad\quad\quad\quad\quad\quad\quad\quad\quad\quad\quad \mathcal{D}{\bm y_{1}}\,\mathcal{D}{\bm y_{0}} 
  \end{align*}
 \\
 This is a $2d$-dimensional integral of the following form
 
\begin{equation}
 p(1|{\bm x}^{\star},D) =\int^{\infty}_{-\infty}\int^{\infty}_{-\infty}  \frac{1}{1 + \mathrm{e}^{a_{1}({\bm y_{1}}-\tilde{{\bm y}}_{1})^{2} -a_{0}({\bm y_{0}}-\tilde{{\bm y}}_{0})^{2}-d\log(C_{01}) + \log(E_{01})}}\,\mathcal{D} {\bm y_{1}} \,\mathcal{D} {\bm y_{0}}
\label{eqn:predictive_distribution_simplified}
 \end{equation}
 with 
 \begin{equation*}
 a_{1} = \frac{1}{2}(\frac{\hat{s}_{1}}{\hat{\psi}_{1}})^{2}\quad a_{0} = \frac{1}{2}(\frac{\hat{s}_{0}}{\hat{\psi}_{0}})^{2}\quad 
\tilde{{\bm y}}_{1} = \frac{{\bm x}^{\star}}{\hat{s}_{1}} - \frac{\hat{s}_{1}}{\hat{\psi}^{2}_{1}} \sum^{N}_{i=1}\delta_{\sigma_{i},1}~{\bm x}_{i}\quad \tilde{{\bm y}}_{0} = \frac{{\bm x}^{\star}}{\hat{s}_{0}} - \frac{\hat{s}_{0}}{\hat{\psi}^{2}_{0}} \sum^{N}_{i=1}\delta_{\sigma_{i},0}~{\bm x}_{i}\quad 
C_{01} = \frac{\hat{\psi}_{0}}{\hat{\psi}_{1}}\quad E_{01} = \frac{\bar{\sigma}_{0}}{\bar{\sigma}_{1}}
 \end{equation*}
The MAP estimates for the hyperparameters $\psi_ {\sigma}$ and $\alpha_ {\sigma}$ are found by minimising the log-likelihood in the usual way.
We choose a flat prior for $\psi_ {\sigma}$. The only information we have about $\alpha_{\sigma}$ is that it is the precision of the mean ${\bm \mu}_{\sigma}$, and $\geq 0$. Therefore we can choose an entropic prior \cite{Rodriguez2002, Caticha2004, Neumann2007} for $\alpha_{\sigma}$. This also has the advantage of rounding off the Bayesian hierarchy at hyperparameter level. The optimal hyperparameters used in (\ref{eqn:predictive_distribution_simplified}) are:

\begin{equation*}
\hat{\alpha}_ {\sigma} = 0 \quad \hat{\psi}^{2}_ {\sigma} =
\frac{N_ {\sigma}{\bm \Sigma}_ {\sigma}^{2}}{(N_ {\sigma}-1)d}
\end{equation*}

where 
\begin{equation*}
N_{\sigma} =\sum^{N}_{i=1}\delta_{\sigma_{i}, \sigma}\quad \langle {\bm x}  \rangle_{\sigma} = \frac{1}{N_ {\sigma}} \sum^{N}_{i=1}\delta_{\sigma_{i}, \sigma}~{\bm x}_{i}\quad 
\langle {\bm x}^{2} \rangle_{\sigma} =\frac{1}{N_ {\sigma}} \sum^{N}_{i=1}\delta_{\sigma_{i}, \sigma}~{\bm x}^{2}_{i}\quad {\bm \Sigma}_ {\sigma}^{2} = \langle {\bm x}^{2} \rangle_ {\sigma} - \langle {\bm x} \rangle_ {\sigma}^{2}
\end{equation*}

\subsection*{Evaluating integral $p(1|{\bf x^{\star}},D)$ for finite d}

We rotate the basis for the vectors ${\bm y_{1}}$ and ${\bm y_{0}}$ such that the first unit vector in the space of ${\bm y_{1}}$ ($y_{11}$) points towards $\tilde{{\bm y}}_{1}$, and the first unit vector in the space of ${\bm y_{0}}$ ($y_{01}$) points in  the direction of $\tilde{{\bm y}}_{0}$. This results in a formula in terms of the absolute values $\tilde{y}_{1} = |\tilde{{\bm y}}_{1}|$, and $\tilde{{y_{0}}} = |\tilde{{\bm y_{0}}}|$ alone:
\begin{align}
p(1|{\bm x}^{\star},D) 
&= \int^{\infty}_{-\infty}\int^{\infty}_{-\infty}\int^{\infty}_{-\infty}\int^{\infty}_{-\infty} W(Y_{1})\, W(Y_{0})\frac{1}{1 + \mathrm{e}^{a_{1}Y_{1} - a_{0}Y_{0} + a_{1} (y_{11}-\tilde{y}_{1})^{2} - a_{0} (y_{01}-\tilde{y}_{0})^{2}-d \log(C) - \log(E)}}\nonumber
\\
&\quad\quad\quad\quad\quad\quad\quad\quad\quad\quad\quad\quad\quad\quad\quad\quad\quad\quad\quad\quad\quad\quad\quad\quad\quad\quad\quad\quad\quad\quad\quad\quad
\rmd Y_{1}\, \rmd Y_{0}\,\mathcal{D} y_{11}\,\mathcal{D}y_{01}
\label{enq:4d_prediction_formula}
\\\nonumber
\text{where, for} \,\, d \geq 3,\nonumber
\end{align}
\begin{equation}
W(Y_{\sigma}) = \int^{\infty}_{-\infty} \mathcal{D}{y_{2}}\ldots  \mathcal{D}{y_{d}}~\delta[Y_{\sigma}-\sum^{d}_{i=2} y_{i}^{2}] = \int_{-\infty}^{\infty} \frac{\rmd v}{2\pi}~\frac{\mathrm{e}^{ivY_{\sigma}}}{(1+2iv)^{\frac{d-1}{2}}} = \frac{2^{-\frac{(d-1)}{2}}}{\Gamma{(\frac{d-1}{2})}}~\theta(Y_{\sigma})~~\mathrm{e}^{[-\frac{Y_{\sigma}}{2} ~+~ (\frac{d-3}{2})\log(Y_{\sigma})]} 
\label{eqn:W(Y_sigma)}
\end{equation}
Substituting (\ref{eqn:W(Y_sigma)}) into (\ref{enq:4d_prediction_formula}) gives

\begin{align}
p(1|{\bm x}^{\star},D) &= \frac{2^{-(d-1)}}{\Big(\Gamma{(\frac{d-1}{2})}\Big)^2}~\int^{\infty}_{-\infty} \int^{\infty}_{-\infty} \int^{\infty}_{0} \int^{\infty}_{0} \frac{e^{-\frac{1}{2}(Y_{1}+Y_{0})\,+\,(\frac{d-3}{2})\log(Y_{1}Y_{0})}}{1 +\mathrm{e}^{a_{1}Y_{1} - a_{0}Y_{0} + a_{1} (y_{11}-\tilde{y}_{1})^{2} - a_{0} (y_{01}-\tilde{y}_{0})^{2}-d \log(C) - \log(E)}}\nonumber
\\
&\quad\quad\quad\quad\quad\quad\quad\quad\quad\quad\quad\quad\quad\quad\quad\quad\quad\quad\quad\quad\quad\quad\quad\quad\quad\quad\quad\quad\quad\quad\quad\quad\quad \rmd Y_{1}\, \rmd Y_{0}\,\mathcal{D} y_{11}\,\mathcal{D}y_{01}
\label{eqn:4d_prediction_formula_simplified}
\end{align}\\
where (\ref{eqn:4d_prediction_formula_simplified}) is $4$-dimensional integral. Thus for $d \geq 3$ we have shown that solving the $d$-dimensional Bayesian integrals analytically has simplified our predictive distribution to a $4$-dimensional integral (FD).

\begin{figure}[t]  
\centering
\includegraphics[height=6cm, width=12cm, scale=0.75]{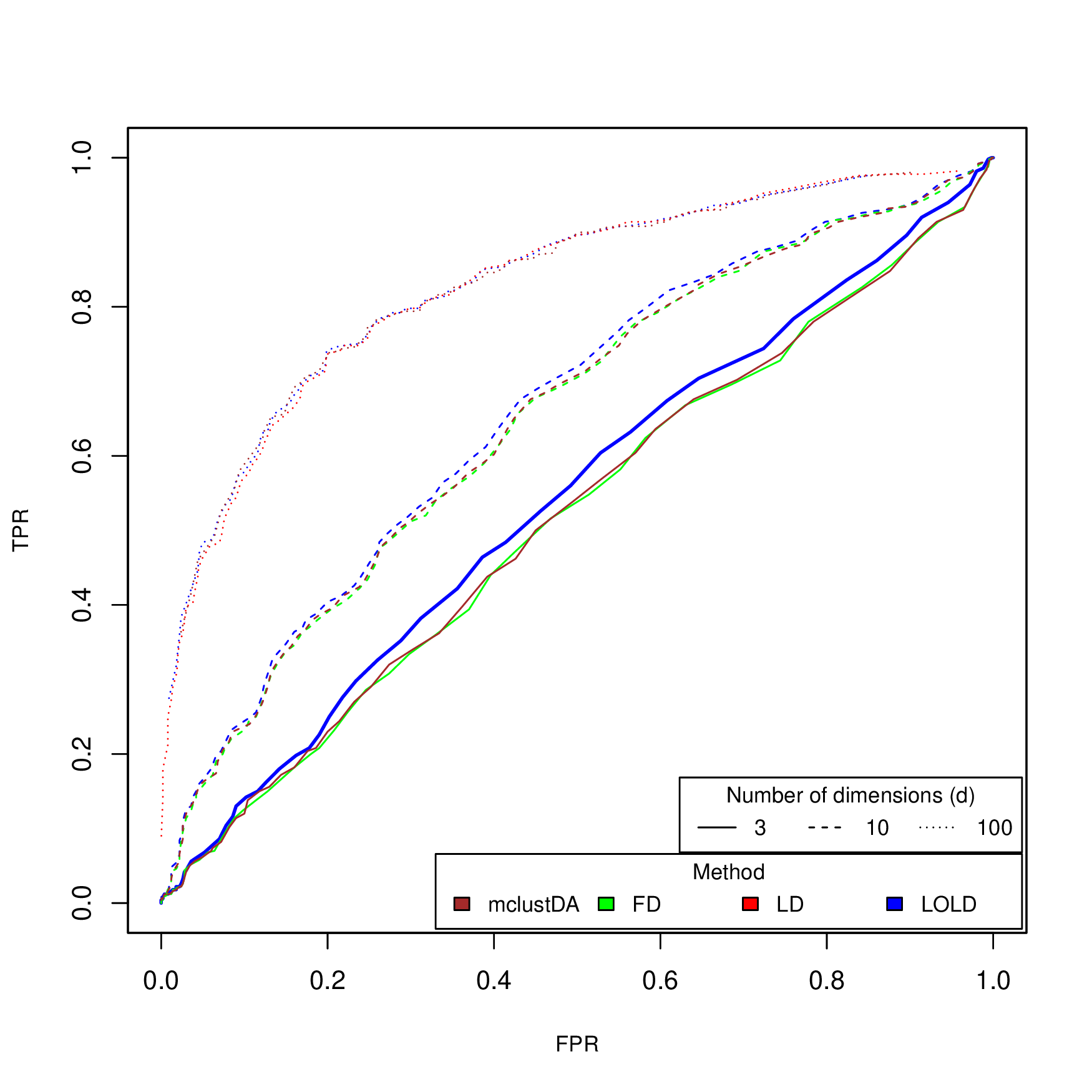}
\caption{ROC curves for the simulated data sets using the different versions of our Bayesian method for $d=3$ (solid line), $d=10$ (dashed line), and $d=100$ (dotted line).}
\label{fig:roc_curves_all_versions_comparison_simulated_c2_data}
\end{figure}

\subsection*{Evaluating integral $p(1|{\bf x^{\star}},D)$ for large d}

For large $d$, we can use the central limit theorem to write $Y_{\sigma} = (d-1) + Y_{\sigma}^{\star}\sqrt{2(d-1)}$, where $Y_{\sigma}^{\star} \sim \mathcal{N}(0,1)$. This gives

\begin{equation}
W(Y_{\sigma}) = \int^{\infty}_{-\infty}  \int^{\infty}_{-\infty} \rmd Y_{\sigma}^{\star}\, \mathcal{D}Y_{\sigma}^{\star}~\delta \Big[Y_{\sigma} - \Big((d-1) + Y_{\sigma}^{\star}\sqrt{2(d-1)}\Big)\Big]
\label{eqn:W(Y_sigma)_central_limit_theorem}
\end{equation}
Substituting (\ref{eqn:W(Y_sigma)_central_limit_theorem}) into (\ref{enq:4d_prediction_formula}) gives

\begin{align}
p(1|{\bm x}^{\star},D) &=  \int^{\infty}_{-\infty}  \int^{\infty}_{-\infty} \int^{\infty}_{-\infty} \frac{1}{1 + \mathrm{e}^{(a_{1}-a_{0})(d-1) + \sqrt{2(d-1)}\sqrt{a_{1}^2+a_{0}^2}w} \, \mathrm{e}^{a_{1}(y_{11}-\tilde{y}_{1})^{2} - a_{0}(y_{01}-\tilde{y}_{0})^{2}-d\log(C) - \log(E)}}\nonumber
\\
&\quad\quad\quad\quad\quad\quad\quad\quad\quad\quad\quad\quad\quad\quad\quad\quad\quad\quad\quad\quad\quad\quad\quad\quad\quad\quad\quad\quad\quad\quad\quad\quad\quad\quad\quad
\mathcal{D} w \,\mathcal{D} y_{11} \,\mathcal{D} y_{01}
\label{eqn:3d_prediction_formula}
\end{align}
where we have combined the sum $a_{1}Y_{1}^{\star}-a_{0}Y_{0}^{\star}$ into a single Gaussian variable $w$, with variance $a_{1}^2+a_{0}^{2}$. Thus for large $d$ we reduce the $4$-dimensional integral to a $3$-dimensional one (LD). 

\begin{figure}[t]  
\centering
\includegraphics[height=6cm, width=12cm, scale=0.75]{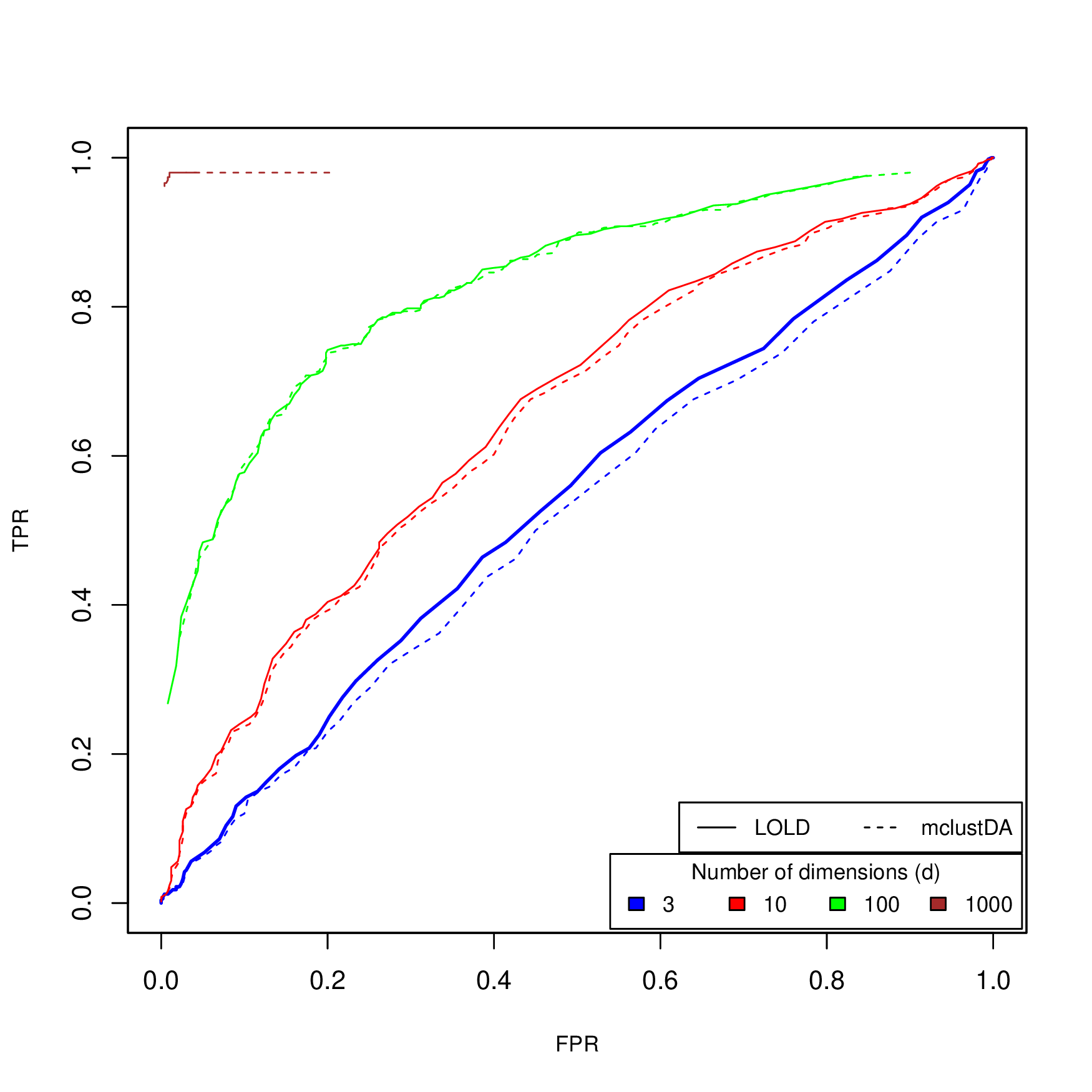}
\caption{ROC curves for the simulated data sets $d=3,10,100,1000$, using LOLD (solid line) and \textit{mclustDA} (dashed line).}
\label{fig:roc_curves_LOLD_mclust_simulated_c2_data}
\end{figure}

\subsection*{Evaluating integral $p(1|{\bf x^{\star}},D)$ for leading order large d}
If we only take the leading order terms (the other terms become negligible as $d$ gets large), then (\ref{eqn:3d_prediction_formula}) becomes a simple formula 

\begin{equation*}
p(1|{\bm x}^{\star},D) = (1+\mathrm{e}^{\,d\,\Lambda_{10}})^{-1} ,\quad\quad \text{with} \quad  \Lambda_{10} = a_{1}-a_{0} - \log(C) + \frac{a_{1}\tilde{y}_{1}^{2}}{d}-\frac{a_{0}\tilde{y}_{0}^{2}}{d}
\end{equation*}
Thus we have 
\begin{equation*}
p(1|{\bm x}^{\star},D)= 
\begin{cases}  0 \,+\, \mathcal{O}(\mathrm{e}^{-d\,\Lambda_{10}}) & \text{if} \quad\Lambda_{10}>0
\\
1 \,-\,\mathcal{O}(\mathrm{e}^{-d\,|\Lambda_{10}|}) & \text{if} \quad \Lambda_{10}<0
\end{cases}
\end{equation*}
which gives
\begin{equation}
p(1|{\bm x}^{\star},D)  = \theta(-\Lambda_{10}) \,\pm\, \mathcal{O}(\mathrm{e}^{-d\,|\Lambda_{10}|})
\label{eqn:simple_predictive_equation}
\end{equation}
Thus we have shown that for leading order large $d$ the $2d$-dimensional predictive distribution simplifies to a simple formula (LOLD) in (\ref{eqn:simple_predictive_equation}).

\section{Simulation study}

In this section, we apply our proposed Bayesian method, and \textit{mclustDA} to simulated data sets.
In order to compare our proposed Bayesian method to \textit{mclustDA}, we simulated data sets consisting of $N=100$ observations, and their corresponding classifications. The model complexity (the number of unknown parameters) was varied by increasing the dimension of each data set, from the lowest dimension that our method can be applied to, $d=3$, to $d=10, 100, 1000,10000$. The last dimension cannot be handled by \textit{mclustDA}, and is used to illustrate the power of our Bayesian method in very large dimensions that are typical of large biomedical data sets \cite{Clarke2008, Michiels2011, Wang2008, Rinaldis2013, Naturebreast2012, Natureovarian2011, CCLE2012}.

\begin{figure}[t] 
\centering
\includegraphics[height=6cm, width=12cm, scale=0.75]{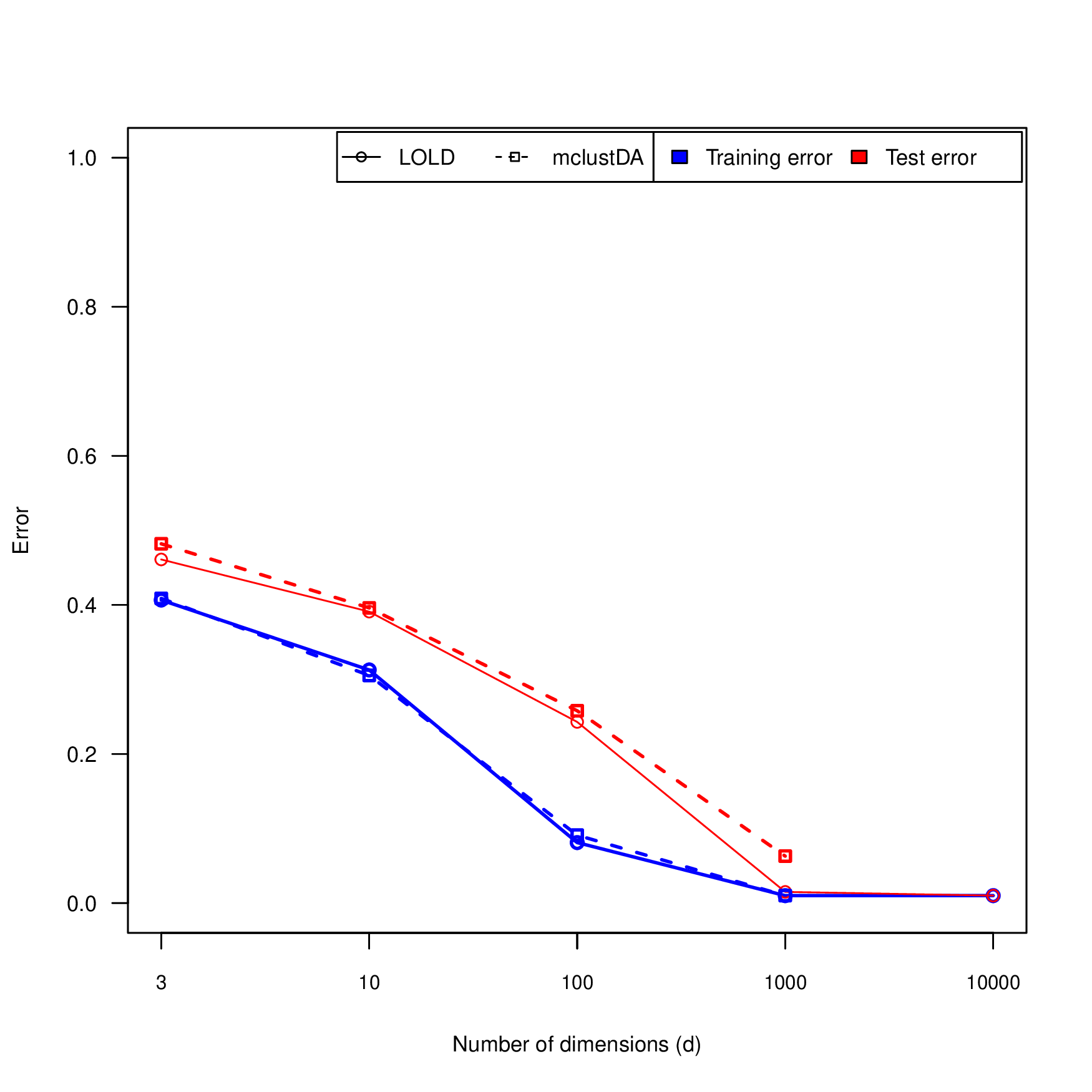}
\caption{Error curves for the simulated data sets with $d=3,10,100,1000, 10000$ at a decision threshold of $0.5$, using LOLD (solid line), and \textit{mclustDA} (dashed line).  The performance improves in high dimensions as the peaks (mass) of the Gaussian distributions move away from each other whilst the width of each distribution remains constant \cite{Bishop2006}.}
\label{fig:error_curves_simulated_c2_data}
\end{figure}
We simulated data sets whose classes had the same mean vector in order to test the predictive and classification power of our methods, since well-separated classes are easier to classify particularly in higher dimensions \cite{Scott1983, Bishop2006}. The variances of the classes were set to be weakly different values, in order to allow one class to nest inside the space of the other. This allowed the classes to become more separated as the dimension increased, since the location of the peaks (mass) of the Gaussian distributions move away from each other whilst the width of each distribution increases much slower (until it becomes constant) as the dimension increases \cite{Bishop2006}. In contrast, where the mean vectors and variances are the same, the classes are indistinguishable \cite{Bishop2006}.

For each dimension, two classes, $\sigma = \{0,1\}$, of observations were simulated according to ${\bm x}|\sigma \sim \mathcal{N}({\bm \mu}_ {\sigma}, \lambda^{2}_ {\sigma} \, \mathbb{I}_{d\times d})$, with $\mu_{0} = \mu_{1} = 0$, $\lambda_{0} = 0.24$, and $\lambda_{1} = 0.28$. The number of observations in each class was kept the same ($N_{0}=N_{1}=50$). Classifications, $\sigma=\{0,1\}$,  were generated for the number of observations in each class respectively. 
The model parameters, class-imbalance parameters, and number of observations for each dimension are summarised in Table \ref{table:simulated_data_sets}.

\begin{table}[h] 
\centering
\caption{Parameters of the simulated isotropic Gaussian data sets consisting of two distinct classes.}
\vspace*{2mm}
\begin{tabular}{|cccccccccc|}  
\toprule
& $N$ &$N_{0}$ &$N_{1}$ &$\bm \mu_{0}$ & $\bm \mu_{1}$ & $\lambda_{0}$ & $\lambda_{1}$ & $\bar{\sigma}_{0}$ &$\bar{\sigma}_{1}$ \\
\midrule
& $100$ &$50$ &$50$ & $0$ & $0$ & $0.24$ & $0.28$ &$0.5$ &$0.5$\\
\bottomrule
\end{tabular}
\label{table:simulated_data_sets}
\end{table}

In order to determine when to use the different versions (FD, LD, LOLD) of our method, we applied all three versions to the simulated data sets with relatively low dimensions ($d=3,10$).  Only LD and LOLD were applied to the simulated data sets with $d=100$. \textit{mclusDA} was applied as a reference.
The results using a threshold of $0.5$ are shown in  Table \ref{table:types_of_bayesian_method_comparison}. The receiver operating characteristic (ROC) curves \cite{Gerds2008} for all the versions and dimensions are displayed in Figure \ref{fig:roc_curves_all_versions_comparison_simulated_c2_data}.
We observe that LOLD preforms best which suggests that the leading order (constant) $\mathcal{O}(d)$ terms dominate the integrated terms which are $\mathcal{O}(1)$. Therefore we continue our analysis using LOLD.

The ROC curves for both methods in the different dimensions are shown in Figure \ref{fig:roc_curves_LOLD_mclust_simulated_c2_data}. As expected, neither LOLD nor \textit{mclustDA} perform well for $d=3$ as they cannot distinguish between the classes. Both methods improve as the dimension of the data set increases given that it lives in subspaces with an effective dimensionality less than $d$ \cite{Bishop2006, Scott1983}, although LOLD outperforms \textit{mclustDA} on average in all dimensions. For $d=10000$, LOLD performs with $99$ per cent accuracy.

Using a classification threshold of $0.5$, Figure \ref{fig:error_curves_simulated_c2_data} illustrates the average training and test error for both methods in different dimensions. We notice that for LOLD overfitting is generally reduced as the number of dimensions increases until it ceases for $d=1000$. For \textit{mclustDA}, overfitting is reduced at a slower rate but remains for $d=1000$. Therefore we have demonstrated that on average, LOLD performs as well as or better than \textit{mclustDA}, and reduces overfitting until it ceases for $d=1000$. 

\begin{table}[h] 
\centering
\caption{Average performance comparison for the different methods.}
\vspace*{2mm}
\begin{tabular}{|c||cccc|} 
\toprule
$d$ &Method &Threshold &Accuracy &Computation time (mins) \\
\midrule
& \textit{mclustDA} &$0.5$ &$0.52$ & $0.01$\\[0.5ex]
\raisebox{-1.5ex}{$3$}  & FD &$0.5$ &$0.52$ & 32551.78\\[0.5ex]
& LD &$0.5$ &$0.52$ & 2214.55\\[0.5ex]
& LOLD &$0.5$ &$0.54$ & 0.00\\[1ex]
\hline
\hline
\\[-2ex] 
& \textit{mclustDA}  &$0.5$ &$0.60$ & $0.01$\\[0.5ex]
\raisebox{-1.5ex}{10} & FD &$0.5$ &$0.61$ & $57295.17$\\[0.5ex] 
& LD &$0.5$ &$0.61$ & $3897.88$\\[0.5ex]
& LOLD &$0.5$ &$0.61$ & $0.00$\\[1ex]
\hline
\hline
\\[-2ex] 
& \textit{mclustDA}  &$0.5$ &$0.74$ & $0.09$\\[0.5ex]
\raisebox{-0.5ex}{100} 
& LD &$0.5$ &$0.76$ & $4263.10$\\[0.5ex]
& LOLD &$0.5$ &$0.76$ & $0.00$
 \\[1ex]
 \hline
\hline
\\[-2ex] 
& \textit{mclustDA}  &$0.5$ &$0.94$ & $9.14$\\[0.5ex]
\raisebox{1ex}{1000} 
& LOLD &$0.5$ &$0.99$ & $0.00$
 \\[1ex]
  \hline
\hline
\\[-2ex] 
{10000}
& LOLD &$0.5$ &$0.99$ & $0.07$
 \\[1ex]
\bottomrule
\end{tabular}
\label{table:types_of_bayesian_method_comparison}
\end{table}

\section{Application to triple-negative breast cancer data}
In this section, we apply our proposed  Bayesian method, and \textit{mclustDA} to a real example. The gene expression data set for triple-negative breast cancer patients (\textit{exprset2}) from \textit{de Rinaldis et al.} \cite{Rinaldis2013} has been used. 
The data set consists of $N=176$ patients, 
who were treated at Guy's and St Thomas's Hospitals, London, UK between 1979 and 2001, and had at least $5.5$ years follow-up. Each patient had $d=22035$ gene expressions recorded.  Patients with missing data or who were lost to follow up were excluded from our analysis. This yielded $N=165$ patients to study.

Patients who survived for an interval of at least $5$ years from initial diagnosis were designated as class $\sigma=1$, the good prognosis group: GP($1$). Patients who died from breast cancer within $5$ years were designated as class $\sigma=0$, the poor prognosis group: GP($0$). The aim was to use the patients' gene expression data to predict which prognosis group they belong to. The total number of patients, the number of patients in each class, and label-imbalance parameters are summarised in Table \ref{table:exprset2_data_set}.

\begin{table}[h] 
\centering
\caption{\textit{exprset2} data set.}
\vspace*{2mm}
\begin{tabular}{|cccccccc|}  
\toprule
$d$ &GP($0$) &GP($1$) &$N$ &$N_{0}$ &$N_{1}$ & $\bar{\sigma}_{0}$ &$\bar{\sigma}_{1}$ \\
\midrule
$22035$ & $<5 yrs$ & $\geq 5 yrs$ &$165$ &$41$ & $124$ &$0.25$ &$0.75$\\
\bottomrule
\end{tabular}
\label{table:exprset2_data_set}
\end{table}

In order to apply \textit{mclustDA} to the \textit{exprset2} data set, we  reduced the number of dimensions a priori since \textit{mclustDA} could not handle such large dimensions. Based on the hypothesis that genes with greater correlation with outcome are likely candidates for reporting prognosis \cite{Veer2002, Vijver2002}, the correlation coefficient of the expression for each gene, and outcome was calculated. The genes were then ranked-ordered according to the magnitude of the correlation coefficient \cite{Veer2002, Vijver2002}. The highest ranked $100$, $1000$ and $10000$ genes were chosen. \textit{mclustDA}  was applied to the first two reduced \textit{exprset2} data sets consisting of $N=165$ patients and $d=100$ genes, and  $N=165$ patients and $d=1000$ genes respectively. LOLD was applied to all three reduced data sets, and the full data set. The results using a decision threshold of $0.5$ are shown in Table \ref{table:exprset2_data_results}. LOLD performed as well as \textit{mclustDA}, and was more efficient for $d=100$. For $d=1000$, LOLD was both more accurate and efficient than \textit{mclustDA}. In higher dimensions, overfitting causes the predictive accuracy of LOLD to decrease. Using a classification threshold of $0.5$, Figure \ref{fig:error_curves_exprset2_c2_data} illustrates the training, and test error for both methods. Compared to \textit{mclustDA}, LOLD reduces overfititng for the two lower dimensions. For the two higher dimensions, LOLD is clearly overfitting although the optimal predictive accuracy only decreases by 6 per cent.

\begin{table}[h] 
\centering
\caption{\textit{exprset2} data set results.}
\vspace*{2mm}
\begin{tabular}{|c||cccc|} 
\toprule
Method &$d$ &Threshold &Accuracy &Computation time (mins) \\
\midrule
& $100$ &$0.5$ &$0.76$ & $0.09$\\[-1ex]
\raisebox{1.5ex}{\textit{mclustDA}}  & $1000$ &$0.5$ &$0.75$ & $11.36$\\[2ex]
\hline
\hline
\\[-2ex] 
& $100$ &$0.5$ &$0.76$ & $0.00$\\[0.5ex]
\raisebox{-1.5ex}{LOLD} & $1000$ &$0.5$ &$0.76$ & $0.03$\\[0.5ex] 
& $10000$ &$0.5$ &$0.74$ & 0.40\\[0.5ex]
& $22035$ &$0.5$ &$0.70$ & $1.67$
\\[1ex]
\bottomrule
\end{tabular}
\label{table:exprset2_data_results}
\end{table}

 Additionally, we now know a posteriori that the predictive information lies in the first $1000$ rank-ordered genes for this data set. This would not have been possible with \textit{mclustDA} as it was less accurate for these $1000$ genes, and could not handle the first $10000$ rank-ordered genes. This information may have been lost using a dimension reduction technique a priori, since it would require a user-defined arbitrary cut-off which may have been less than the first $1000$ rank-ordered genes.
\begin{figure}[t]  
\centering
\includegraphics[height=6cm, width=12cm, scale=0.75]{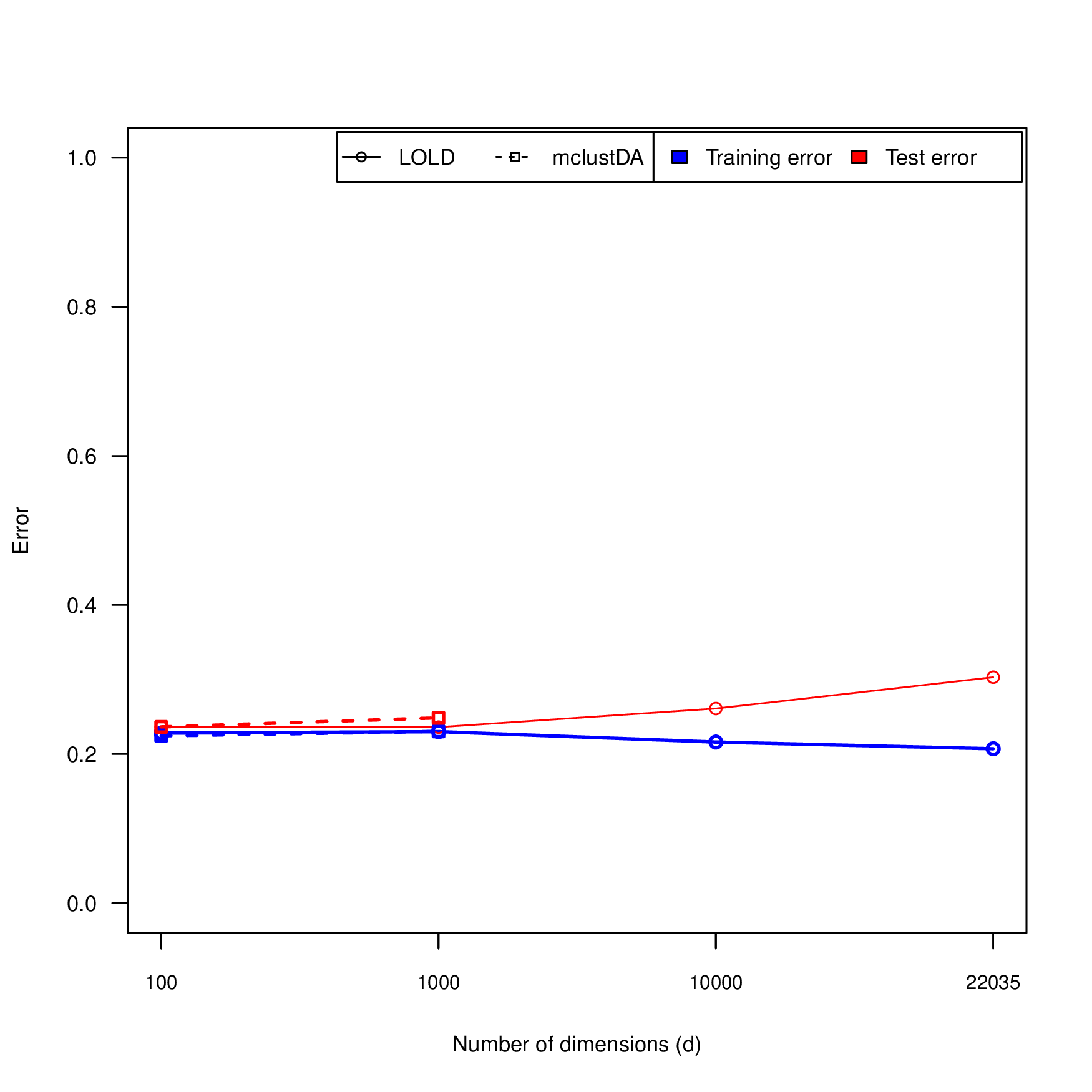}
\caption{Error curves for the exprset2 data sets $d=100,1000,10000, 22035$ at a decision threshold of $0.5$, using LOLD (solid line), and \textit{mclustDA} (dashed line).}
\label{fig:error_curves_exprset2_c2_data}
\end{figure}

\section{Discussion}

We have proposed a Bayesian binary outcome prediction method that can be applied to any arbitrary number of dimensions without any approximations at parameter level. Our method overcomes the computational inability associated with other methods in large dimensions without sacrificing any information from the data set. Although three prediction formulas, (FD, LD, LOLD) have been developed, we have shown that it suffices to use the most efficient version, LOLD, for any arbitrary dimension, without compromising predictive performance. 
We compared our method to \textit{mclusDA} \cite{Fraley2002, Mclustversion4} due to its Bayesian approach, and  wide application \cite{Dean2006, Fraley2007, Iverson2009, Murphy2010, Andrews2012}.  For data sets where the methods produce the same results, our method is more efficient. For high-dimensional data sets, our Bayesian method is superior in both accuracy and efficiency. Our method also succeeds in reducing overfitting although it remained in very high-dimensions.

Our method can be seen as an extension for \textit{mclustDA}, in a homogeneous binary class setting, where we take the uncertainty at parameter level into account instead of estimating the Bayesian integrals using EM. However the methods differ in  their application to the training set. We use a supervised method by applying our Bayesian method to training set using both the training data and outcome information whilst \textit{mclustDA} applies an unsupervised model-based clustering algorithm to the training data only. A limitation of our approach is that we must set the number of classes in the data set, and select the optimal model a priori, whilst \textit{mclustDA} calculates this from the data. 
We also limited our approach to only two homogeneous classes. A natural extension would be to expand our approach to consider more than two classes. Our method could also extend to heterogeneous classes. The would entail fitting a more complex GMM to each class. We would also need to consider Bayesian model selection in order to choose the optimal parametrisation for each GMM. The uncertainty in model selection, and at hyperparameter level could also be incorporated into our frame work. However the risk of heavily overfitting remains for all these ideas, as the number of unknown parameters, and hyperparameters will increase considerably. 

Finally, we have only compared our method to the \textit{mclustDA}. Many other model-based discriminant analysis methods exist that overcome computational inability in high-dimension \cite{McNicholas2011, Bouveyron2014}. These include using variable selection \cite{Raftery2006, Murphy2010}, and combining subspace clustering with constrained and parsimonious models \cite{Bouveyron2007, Bouveyron2014}. However, these methods may still suffer from loss of information, and/or underestimated uncertainty at parameter level. Overfitting remains where the imbalance between the number of observations and dimensions remains high, even after reducing the dimensionality of the data set. The method we propose overcomes these problems by reducing the number of dimensions analytically, whilst also taking the uncertainty at parameter level into account. No information is lost during this process, and we postpone the use of computer resources until the resulting formulae can no longer be simplified.

\newpage
\bibliographystyle{wileyj}
\bibliography{refs.bib}
\end{document}